\documentstyle[12pt]{article}
\begin{document}
\renewcommand{\theequation}{\thesection.\arabic{equation}}
\newcommand{\be}{\begin{equation}}
\newcommand{\ee}{\end{equation}}
\begin{flushright}
BUTP-95/4
\end{flushright}
\vskip 1.5cm
\begin{center}
{\Large\bf
Chiral symmetry and spectrum of Euclidean Dirac operator in QCD
}
\footnote{Talk given at the International Conference on {\it Chiral Dynamics in
Hadrons and Nuclei}, Seoul, February 1995.}\\
\vskip 1.0cm

{\bf A.V. Smilga}\\
{\em Institute for Theoretical Physics, University of Bern, Sidlerstrasse 5,
Bern CH-3012, Switzerland}
\footnote{Permanent adress: Institute for Theoretical and Experimental Physics,
B. Cheremushkinskaya 25, Moscow 117259, Russia.}\\

\vskip 1cm

March 1995
\end{center}

\vskip 1.5cm

\begin{abstract}
Some exact relations for the spectral density $\rho(\lambda)$ of the Euclidean
Dirac operator in $QCD$ are derived. They follow directly from the
chiral symmetry of the $QCD$ lagrangian with massless quarks. New results
are obtained both in thermodynamic limit when the Euclidean volume $V$
is sent to infinity and also in the theory defined in finite volume where
the spectrum is discrete and  a nontrivial information on $\rho(\lambda)$
in the region $\lambda \sim 1/(|<\bar{q}q>_0|V)$ (the characteristic level
spacing) can be obtained. These exact results should be confronted with
"experimental" numerical simulations on the lattices and in some particular
models for $QCD$ vacuum structure and may serve as a nontrivial test of the
validity of these simulations.
\end{abstract}

\section{Introduction}
The notion of spectral density of the Euclidean Dirac operator in $QCD$
has been brought into discussion some years ago in the pioneering paper
by Banks and Casher \cite{Banks}. They have got the famous formula
 \be
 \label{Banks}
\rho(0) = - \frac 1\pi <\bar{q}q>_0
 \ee
relating the infrared limit of the averaged spectral density with the quark
condensate --- the order parameter for the spontaneous chiral symmetry
breaking. $\rho(\lambda)$ is defined as
  \be
  \label{rodef}
\rho(\lambda) = \frac 1V <\omega(\lambda, A)>_A
  \ee
where
  \be
  \label{omega}
\omega(\lambda, A) = \sum_n \delta(\lambda - \lambda_n)
  \ee
is the microscopic spectral density of the Dirac operator in a given background
field A in a finite Euclidean volume $V$. The averaging is taken over all
gluon fields with the standard Yang-Mills measure involving the determinant
factor for $N_f$ fermions with the common mass $m$. The combination
 $V\rho(\lambda)d\lambda$ defines the average number of eigenvalues
 of Dirac operator in the interval $(\lambda, \lambda + d\lambda)$.

 Let us recall how the Banks-Casher relation (\ref{Banks})
(which holds in the thermodynamic limit $V \rightarrow \infty$, with the mass
$m$ being kept small but fixed)
is derived.
Treating the gauge field $A_\mu^a(x)$ as an external field, the fermion
Green's function is given by
  \be
  \label{Green}
S_A(x,y) = <q(x) \bar{q}(y)>_A = \sum_n \frac{u_n(x) u_n^\dagger(y)}
{m - i\lambda_n}
  \ee
where $u_n(x)$ and $\lambda_n$ are eigenfunctions and eigenvalues of
 the Euclidean Dirac operator:
  \be
  \label{Dirac}
\not\!\!{\cal D} u_n(x) = \lambda_n u_n(x)
  \ee
Note that the spectrum of the theory enjoys the chiral symmetry:
 \be
 \label{spesym}
\psi_n \rightarrow \gamma^5 \psi_n,  \lambda_n \rightarrow -\lambda_n
 \ee
And, except for zero modes, the eigenfunctions occur in pairs with opposite
eigenvalues. Setting $x=y$ and integrating over $x$, the representation
(\ref{Green}) therefore implies
 \be
 \label{condsum}
 \frac 1V \int_V dx <q(x) \bar{q}(y)>_A = - \frac {2m}V \sum_{\lambda_n > 0}
\frac 1{m^2 + \lambda_n^2}
 \ee
where the zero mode contributions have been dropped out (it is justified in
 the thermodynamic limit.
See Ref.\cite{Leut} for details). In the limit $V \rightarrow \infty$,
 the level spectrum becomes dense and we can trade the sum in the r.h.s.
of Eq.(\ref{condsum}) for the integral. We get
  \be
  \label{condint}
<\bar{q}q>_0 = - 2m \int_0^\infty d\lambda \frac {\rho(\lambda)}
{m^2 + \lambda^2}
  \ee

The integral (\ref{condint}) diverges at large $\lambda$. For $\lambda
\gg \Lambda_{QCD}$ the spectral density is not sensitive to gluon
 vacuum fluctuations and behaves in the same way as for free fermions:
  \be
  \label{free}
\rho^{free}(\lambda) \sim \frac {N_c \lambda^3}{4\pi^2}
  \ee
This perturbative quadratically divergent piece in $<\bar{q}q>_0$
 is proportional to the quark mass and is related just to the fact that mass
terms in the lagrangian break the chiral symmetry explicitly. To get the
truly non-perturbative quark condensate which is the order parameter
of the {\it spontaneous} chiral symmetry breaking, this perturbative
divergent part should be subtracted. As a result, the quark condensate is
 related to the region of small $\lambda$: $\lambda \sim m \ll \Lambda_{QCD}$.
It is not difficult to see that the finite mass-independent contribution
appears if $\rho(0) \neq 0$ and the relation (\ref{Banks}) holds.

\section{Thermodynamic limit: $\lambda \neq 0$.}
\setcounter{equation}0
In this section, we derive a new formula describing the behavior of the
spectral density $\rho(\lambda)$ in the region $m \ll \lambda \ll
\Lambda_{QCD}$ in the thermodynamic limit $V \rightarrow \infty$
\cite{Stern}.
 To this end, let us consider the 2-point correlator

 \be
   \label{Kab}
K^{ab} = \int d^4x \int d^4y <0|S^a(x) S^b(y)|0>
   \ee
with $S^a =\bar{q} t^a q$ where $t^a$ is the generator of the {\it flavour}
$SU(N_f)$ group. The correlator $K^{ab}$ can be evaluated in the same way
as $<\bar{q} q>_0$. First, fix a particular gluon background. As
$Tr\{t^a\} = 0$, only the connected part depicted in Fig.1a contributes, and
one obtains
   \be
   \label{KabA}
K_A^{ab} = -\int d^4x \int d^4y {\rm Tr} \{t^a {\cal G}_A(x,y) t^b
{\cal G}_A(y,x) \}
   \ee
Substituting here the spectral decomposition for the Green's function
(\ref{Green}) and taking into account the chiral symmetry of the spectrum
(\ref{spesym}), we get
   \be
   \label{Kabsum}
K_A^{ab} = - \delta^{ab} \sum_{\lambda_n>0} \frac{m^2 -\lambda_n^2}
{(m^2+\lambda_n^2)^2}
  \ee
Rewriting the sum as an integral as in Eq.(\ref{condint}) and averaging
over the gluon background, one gets
    \be
   \label{Kabint}
\frac 1V K^{ab} = - \delta^{ab} \int_0^\infty \frac{\rho(\lambda)(m^2
-\lambda^2)}
{(m^2+\lambda^2)^2} d\lambda
  \ee
On the other hand, the correlator (\ref{Kab}) of the colorless scalar
currents can be evaluated by inserting the complete set of physical
states. For small $m$, i.e. close to the chiral limit, a distinguished
position among the latter belongs to the Goldstone states which appear
due to spontaneous breaking of the chiral symmetry of the QCD lagrangian.
Low energy properties of Goldstones are fixed by chiral symmetry, and
some exact calculations are possible. The chiral lagrangian has the form
\cite{CPT}
  \be
  \label{Lchi}
{\cal L} = F_\pi^2 \left[ \frac 14 {\rm Tr} \{ (\partial_\mu U^\dagger)
(\partial_\mu U) \} + B\  {\rm Re\  Tr}\{{\cal M}U^\dagger\} \right]
\nonumber \\
+ {\rm higher\  order\   terms}
  \ee
where $B = -<\bar{q} q>_0 F_\pi^{-2},\ \  U = \exp \{2i\phi^at^a /F_\pi\}$,
and ${\cal M}$ is the quark mass matrix. If ${\cal M} =
{\rm diag}(m,\ldots,m)$,
the lagrangian (\ref{Lchi}) describes $N_f^2 - 1$ (quasi-) Goldstone states
with the common mass
  \be
  \label{Mpi}
M_\phi^2 = 2mB
  \ee
Consider now the graph in Fig.1b with 2-goldstone intermediate state
 contributing to the correlator (\ref{Kabint}) (obviously, one-goldstone
state does not contribute since \\ $<0|S^a|\phi^b> = 0$). To calculate it, we
need to know the vertex $<0|S^a|\phi^b\phi^c>$. It can be easily found from
the generating functional involving external scalar sources. The latter
is obtained substituting in the effective lagrangian (\ref{Lchi}) the mass
matrix ${\cal M}$ by ${\cal M} + u^at^a$ where $u^a$ is the source for the
scalar current $S^a$. In this way, one gets
 \be
 \label{dabc}
<0|S^a|\phi^b\phi^c> = Bd^{abc}
 \ee
It is very important that the vertex is nonzero only for three or more
flavors. Now we can calculate the graph in Fig.1b absorbing its ultraviolet
divergence into local counterterms contained in higher-order terms in
 the effective lagrangian (\ref{Lchi}) \cite{CPT}. The result reads
  \be
  \label{Kres}
\frac 1V K^{ab} = - \frac{B^2 (N_f^2 - 4)}{32\pi^2 N_f} \delta^{ab}
\ln (M_\phi^2/\mu_{hadr}^2)
  \ee
For massless goldstones, the graph in Fig.1b does exhibit a logarithmic
infrared singularity reflected in the factor $\ln M_\phi^2$ on the r.h.s.
of Eq.(\ref{Kres}). The same circumstance makes our calculation
 self-consistent: since infrared singularity arises from the
 low momenta region, higher derivative terms in Eq.(\ref{Lchi}) can be
neglected.

Now, let us compare Eq.(\ref{Kabint}) with Eq.(\ref{Kres}). Note first
of all that, in contrast to the integral in Eq.(\ref{condint}), the
 constant part $\rho(0)$ does not contribute here at all:
  \be
  \label{int0}
\int_0^\infty \frac{\rho(0)(m^2 -\lambda^2)}
{(m^2+\lambda^2)^2} d\lambda = 0
  \ee
Thus, only the difference $\rho(\lambda) - \rho(0)$ is relevant. It is easy
to see that, in order to reproduce the singularity
$\propto \ln\ M_\phi^2 \propto \ln\ m$, one should have
  \be
  \label{lamb}
 \rho(\lambda) - \rho(0) = C\lambda
  \ee
at small $\lambda$. Comparing the coefficients at the logarithms $\ln m$
and $\ln M_\phi^2$ in Eq.(\ref{Kabint}) and Eq.(\ref{Kres}), we arrive
 at the result
   \be
   \label{rores}
\rho(\lambda) = - \frac 1\pi <\bar{q}q>_0 + \frac {<\bar{q}q>_0^2(N_f^2 -4)}
{32\pi^2 N_f F_\pi^4} |\lambda| + o(\lambda)
   \ee
The structure $|\lambda|$ appears due to the relation $\rho(-\lambda)
= \rho(\lambda)$ being implied by the chiral symmetry (\ref{spesym}). Note
again that this non-analyticity does not appear at $N_f = 2$.

Recently, the preliminary, not yet published data for
the spectral  density $\rho(\lambda)$ calculated on the
lattice with two dynamical fermion flavours
appeared \cite{Christ} . The same quantity has been also evaluated  in the
instanton-antiinstanton liquid model for QCD vacuum configurations
\cite{JacPol}. Results of these two numerical studies are completely
different. Lattice calculation failed to reproduce the exact QCD relation
(\ref{rores}) (they exhibit a large nonzero slope for $\rho(\lambda)$
which should be absent for $N_f = 2$).
 Probably, it is due to the fact that , in the region of
 lattice parameters used in \cite{Christ}, the thermodynamic limit was not
reached yet and the finite volume effects were still important \cite{Norman}.
Obviously, further studies in this direction are highly desirable.

On the other hand, the calculations in the instanton model agree well with
theoretical predictions. The results for the spectral density obtained
in \cite{JacPol} for different number of light quark flavors are presented
in Fig.2. Before comparing them with the theory, two remarks are in order.
\begin{enumerate}
  \item Only non-perturbative instanton-driven part of spectral density has
been determined in \cite{JacPol}. So, the perturbative effects which give
the dominant contribution to the spectral density at large $\lambda$ were
not taken into account, and the comparison with theory makes sense only in
the small $\lambda$ region.
  \item The abrubt falloff and vanishing of $\rho(\lambda)$ at $\lambda = 0$
as measured in \cite{JacPol} is the finite volume effect. The comparison
should be done in the region of not yet too small $\lambda$.
\end{enumerate}

{}From the graphs in Fig.2, one sees that, for $N_f = 2$, the spectral density
is practically flat, while for $N_f = 3$ it rises with a nonzero slope. This is
exactly what the formula (\ref{rores}) requires.

The derivation of the result (\ref{rores}) assumed the spontaneous breaking
of chiral symmetry and the existence of the Goldstone bosons. Thus, it refers
only to the case $N_f \geq 2$. For $N_f = 1$, there is no theoretical result,
and no comparison can be done. It is amusing, however, that, if trying to
"continue analytically" the formula (\ref{rores}) down to the point $N_f = 1$,
one would get a negative slope for $\rho(\lambda)$ which agrees again with
the instanton simulations.

Qualitatively, the different $\lambda$ - dependence
for different $N_f$ is natural. Mean spectral density $\rho(\lambda)$
 is obtained after averaging of the microscopic spectral density
(\ref{omega}) over gluon fields with the weight which involves also the
quark determinant
  \be
   \label{detNf}
[{\rm det} (i\not\!\!{\cal D})]^{N_f} \sim \left[\prod_{\lambda_n > 0}
\lambda_n^2
\right]^{N_f}
  \ee
The larger $N_f$ is, the more the region of small $\lambda$ is suppressed.
There are good reasons to expect that, at $N_f = 0$, with no suppression
at all, quark condensate $\equiv \rho(0)$ is infinite in the thermodynamic
limit \cite{SMvac}.
\footnote{The instanton calculations of Ref.\cite{JacPol} do not exhibit
the infinite quark condensate for the quenched theory, but only a much more
dramatic rise of the spectral density in the low $\lambda$ region. But,
probably, the finite value for $\rho(0)$  is, again, a finite-volume
effect.}

\section{Finite volume: the partition function.}
\setcounter{equation}0
In the rest of this talk, I derive and discuss some relations for
the spectral density in the region of very small $\lambda \sim
1/|<\bar{q}q>|V$. These relations refer not to physical QCD in the infinite
volume, but to the theory defined in the finite Euclidean box. But, as our
main goal is to discover islands of firm theoretical ground in the foreboding
sea of numerical simulations , and the latter are done exclusively in finite
volume, these intrinsically finite volume results can serve this purpose
in exactly the same way as the result (\ref{rores}) derived in the infinite
volume.

But before proceeding to the Dirac operator spectrum, we are in a position
to study the more basic quantity  --- namely, the $QCD$ partition function
at finite volume. Specifically, we will be interested with the dependence
of the partition function on the quark mass matrix ${\cal M}$ and the
vacuum angle $\theta$.

Consider first the theory with only one light quark flavor. It involves
a gap in the spectrum ($U(1)$ axial symmetry is broken not spontaneously
but explicitly by anomaly, and no goldstones appear). Thus, the extensive
property for the partition function holds:
  \be
  \label{Zext}
Z \sim \exp\{-\epsilon_{vac}(m,\theta)V\}
  \ee
If $L \gg \Lambda_{QCD}^{-1}$, the finite volume effects in $\epsilon_{vac}$
are exponentially small \cite{Lee}. Ward identities dictate that
$\epsilon_{vac}$ can
depend on $m$ and $\theta$ not in an arbitrary way, but only as a function
of a particular combination $me^{i\theta}$. Expanding
$\epsilon_{vac}(me^{i\theta})$ in Taylor series (that makes sense as long as
$m \ll \Lambda_{QCD}$) and bearing in mind that
$\epsilon_{vac}$ is real, we get
  \be
  \label{Zm1}
Z \sim e^{\Sigma V m \cos \theta \ + \ O(m^2)}
  \ee
The parameter $\Sigma$ is nothing else but the quark condensate (up to a sign).
Really,
  \be
  \label{sigma}
<\bar{q}q>_0 = - \frac d{dm} \ln Z = -\Sigma
  \ee

Consider now the same problem but for several quark flavors. Then spontaneous
breaking of chiral symmetry occurs, goldstones appear, there is no gap in the
spectrum , and the extensive property (\ref{Zext}) does not hold anymore.
More exactly, if quark masses are non-zero, the property (\ref{Zext}) still
holds when the length of the box $L$ is much larger than the Compton wavelength
of Goldstone particles $\sim 1/m_\phi \sim 1/\sqrt{m\Lambda_{QCD}}$. We will
be interested, however, with the intermediate region
  \be
  \label{region}
\Lambda_{QCD}^{-1} \ll L \ll \frac 1 {\sqrt{m\Lambda_{QCD}}}
  \ee
where the effects due to goldstones on the volume and mass dependence of the
partition function are crucial.

Fortunately, the goldstone properties {\it at low energies} (and, in the
region $L \gg \Lambda_{QCD}^{-1} $, only the low-energy properties of the
Goldstone fields are relevant) are well known. They are described by
 the effective lagrangian (\ref{Lchi}) where, at nonzero $\theta$ ,
${\cal M}$ should be substituted by ${\cal M}e^{i\theta/N_f}$ (again, the
occurence of this particular combination is dictated by Ward identities).
The partition function is given by the functional integral
  \be
  \label{ZU}
Z \sim \int [dU] \exp \left\{ - F_\pi^2 \int_V d^4x  \left[ \frac 14 {\rm Tr}
\{ (\partial_\mu U^\dagger)
(\partial_\mu U) \} + B\  {\rm Re\  Tr}\{{\cal M} e^{i\theta/N_f}
U^\dagger \right] \right\}
  \ee

The crucial observation is that, as long as $L \ll \frac 1{m_\phi} \sim
\frac 1{\sqrt{m\Lambda_{QCD}}}$, only zero Fourier harmonics $U_0$ of the
field $U(x)$ is relevant, and the contribution of the higher harmonics
in the integral (\ref{ZU}) is suppressed. Then the functional integral
is transformed into the finite-dimensional integral
  \be
  \label{ZU0}
Z \sim \int_{SU(N_f)} d\mu(U_0) \exp \{V\Sigma \ {\rm Re\ Tr}
\{{\cal M}e^{i\theta/N_f}U_0^\dagger\}\}
  \ee
The integral extends over the group $SU(N_f)$, and $d\mu(U_0)$ is the
corresponding Haar measure. For $N_f = 2$, the integral (\ref{ZU0}) can be
done explicitly. For $N_f \geq 3$, it is not possible to do analytically in
the general case. Sometimes (for the case ${\cal M} = m\hat{1}$), certain
beautiful results can be obtained \cite{Leut}, but, to derive the {\it sum
rules} for the small eigenvalues of the Dirac operator we are after in
 this talk, the closed analytic expression for $Z(V, {\cal M}, \theta)$
is not actually required.

  The partition function (\ref{ZU0}) is a periodic function of $\theta$
with the period $2\pi$. Thus, it can be presented as a Fourier series
  \be
  \label{ZFour}
Z(\theta)  = \sum_{\nu = -\infty}^{\infty} Z_\nu e^{i\nu\theta}
  \ee
so that
  \be
  \label{Znu}
Z_\nu = \frac 1{2\pi} \int_0^{2\pi} d\theta e^{-i\nu\theta} Z(\theta)
d\theta \nonumber \\
\sim \int_{U(N_f)} d\mu(\tilde{U}) \ ({\rm det}\
\tilde{U})^\nu \exp \{V\Sigma\ {\rm Re\ Tr}
\{{\cal M}\tilde{U}^\dagger\}\}
 \ee
with $\tilde{U} = U_0e^{-i\theta/N_f}$.

\section{Finite volume: sum rules.}
\setcounter{equation}0

The main idea is basically the same as in Sect.2 --- to compare the expression
(\ref{Znu}) derived in the chiral theory with the quark-gluon representation
of the same quantity. $Z_\nu$ may be writen as a functional integral over the
quark and gluon fields restricted on the topological class with a given winding
number
  \be
  \label{nu}
\nu = \frac 1{32\pi^2} \int d^4x \ G^a_{\mu\nu} \tilde{G}^a_{\mu\nu}
  \ee
The integral over the Fermi fields produces the determinant of the Dirac
operator, and we get for $\nu > 0$
  \be
  \label{Znuq}
Z_\nu = \int [dA] e^{- \frac 14 \int d^4x G_{\mu\nu}^a G_{\mu\nu}^a}
(det_f {\cal M})^\nu \prod_{\lambda_n > 0}
det_f (\lambda_n^2 + {\cal M} {\cal M}^\dagger)
  \ee
where the factor $(det_f {\cal M})^\nu$ arises due to the fermion zero modes
$\lambda_n = 0$ which appear on a topologically non-trivial background due to
the index theorem. (If $\nu$ is negative, the factor $(det_f {\cal M})^\nu$
is to be replaced by $(det_f {\cal M}^\dagger)^{-\nu}$ ).

Let us expand now Eqs.(\ref{Znu}) and (\ref{Znuq}) in quark mass and compare
the coefficients of $(det_f {\cal M})^\nu {\rm Tr} \{{\cal M} {\cal M}^\dagger
\}$. On the chiral side of the equality, we  get some group integral which
can be done explicitly. On the quark and gluon side, we get the expression
involving $< \sum_{\lambda_n > 0} 1/\lambda_n^2>_\nu$ where the average
 is defined as
  \be
  \label{aver}
<f>_\nu = \frac {\int [dA] f\ e^{- \frac 14 \int d^4x G_{\mu\nu}^a
G_{\mu\nu}^a}
\left(\prod_{\lambda_n > 0}\lambda_n^2\right)^{N_f} }
{\int [dA]  e^{- \frac 14 \int d^4x G_{\mu\nu}^a G_{\mu\nu}^a}
\left(\prod_{\lambda_n > 0}\lambda_n^2\right)^{N_f}}
  \ee
where the path integral is done over all gauge fields with the topological
charge $\nu$.
Skipping the technical details, we present the final result for the simplest
sum rule thus derived:
  \be
  \label{sum2}
\left< \sum_{\lambda_n > 0} \frac 1{\lambda_n^2} \right>_\nu =
\frac {V^2\Sigma^2}{4k}
  \ee
where $k = |\nu| + N_f$.

Expanding the expressions (\ref{Znu}) and (\ref{Znuq}) further in quark mass
and comparing the coefficients of the two group invariants
$(det_f {\cal M})^\nu ({\rm Tr} \{{\cal M} {\cal M}^\dagger \})^2  $
 and \\$(det_f {\cal M})^\nu {\rm Tr} \{{\cal M} {\cal M}^\dagger
{\cal M} {\cal M}^\dagger\}$ , we get two different sum rules
for the inverse fourth powers of eigenvalues
   \be
   \label{sum4}
\left< \sum_{\lambda_n > 0} \frac 1{\lambda_n^4}\right>_\nu =
\frac {V^4\Sigma^4}
{16k(k^2-1)}; \ \ \ \ \
\left< \left(\sum_{\lambda_n > 0} \frac 1{\lambda_n^2}\right)^2 \right>_\nu =
\frac {V^4\Sigma^4}{16(k^2-1)}
   \ee
(Two different sum rules (\ref{sum4}) are obtained when $N_f \geq 2$.
If $N_f = 1$, there is no difference between
$({\rm Tr} \{{\cal M \cal M}^\dagger \} )^2$
and ${\rm Tr} \{{\cal M \cal M}^\dagger {\cal M \cal M}^\dagger \}$
and only one sum
rule involving inverse fourth powers of $\lambda_n$ for the difference
of the two sums entering Eq.(\ref{sum4}) can be derived).
Further expansion in ${\cal M}$ provides the sum rules with inverse
sixth powers of eigenvalues etc.

Note that the sums (\ref{sum2}) , (\ref{sum4}) are saturated by  few
first eigenvalues. Therefore the sum rules provide the information on
the very bottom of the finite-volume spectrum.

The theoretical results (\ref{sum2}) , (\ref{sum4}) have not yet been
confronted with lattice experiments. However, the numerical calculations
in the instanton
model in the case $\nu = 0$ with different number of quark flavors
 are available. They are in a good agreement with the theory (see Fig.3 and
the discussion thereafter).

\section{Stochastic matrices}
\setcounter{equation}0
 The sum rules (\ref{sum2}) , (\ref{sum4}) are exact theoretical results.
They follow just from the form of $QCD$ lagrangian and the assumption
(which is the experimental fact in $QCD$) that the spontaneous breaking
of chiral symmetry occurs.
It is interesting, however, that much reacher information on the spectrum
of Dirac operator in the region $\lambda \sim 1/(\Sigma V)$ can be obtain
if invoking a certain {\it additional assumption} on the properties of the
functional integral in $QCD$ \cite{Izm}.

Consider the Dirac operator in a particular gauge field background with
the topological charge $\nu$. For simplicity, let us assume that the mass
matrix is diagonal and real. Let us choose a particular basis
 ${\cal E} = \{\psi_n(x)\}$ and present the Dirac operator $-i\not\!\!{\cal D}
+ m_f$ as a matrix in this basis. The particular choice of the basis is not
important for us, but a reader familiar with the model of
 instanton-antiinstanton liquid may think of $\psi_n(x)$ as of zero-mode
solutions for {\it individual} instantons and antiinstantons. What {\it is}
important is that the basis ${\cal E}$ is chosen in such a way that
all states $\psi_n(x)$ have a definite (left or right) chirality. In that
case, the number of left-handed states $n_L$ and the number of right-handed
states $n_R$ in the basis should be related:
  \be
  \label{NMnu}
n_L\ - \ n_R = \ \nu
  \ee
Thereby, the existence of exactly $\nu$ left-handed ($\nu > 0$) or
$-\nu$ right-handed ($\nu < 0$ ) zero eigenvalues of the massless Dirac
operator $-i\not\!\!{\cal D}$ is assured. In this basis, the full Dirac
operator
$-i\not\!\!{\cal D} + m_f$ can be presented as a matrix
   \be
   \label{matr}
-i\not\!\!{\cal D} + m_f \ = \ \left( \begin{array}{cc}
m_f & iT \\ iT^\dagger & m_f \end{array} \right)
   \ee
where $T$ is a general rectangular complex matrix $n_L \times n_R$. We assume
$n_L, n_R$ to be very large and will eventually send them to infinity. In this
limit, the basis ${\cal E}$ spans the whole Hilbert space and the
 representation (\ref{matr}) is just exact.

The partition function involves the Dirac determinant and the integral over
all gluon configurations. If the basis ${\cal E}$ is fixed (the same for all
gluon field configurations), the path integral over the latter can be traded
for an integral over the matrices $T$. Thus, we have
   \be
   \label{ZT}
Z_\nu = \int {\cal D} T P(T) \prod_f^{N_f} det
\left( \begin{array}{cc}
m_f & iT \\ iT^\dagger & m_f \end{array} \right)
   \ee
, the condition (\ref{NMnu}) for the dimensions of the matrix $T$ being
assumed.

And now comes the crucial assumption. Let us assume that the measure $P(T)$ has
the simple form
  \be
  \label{PT}
P(T) = \exp\left\{ -\frac An {\rm Tr} \{TT^\dagger\} + o(TT^\dagger) \right\}
  \ee
The constant $A$ will at the end determine the spectral density $\rho(0)$ in
the thermodynamic limit.

The form (\ref{PT}) of the weight is rather natural. It respects chiral
symmetry, analytic in $T$ , and is conceptially similar to the Gibbs
statistical distribution. However, a rigourous {\it proof} that $P(T)$
{\it should} have this form is absent by now. Therefore, the results for
the microscopic spectral density $\rho(\lambda \sim 1/\Sigma V)$ derived
in (\cite{Izm}) which follow from the representation (\ref{ZT}) for the
partition function and the heuristic assumption (\ref{PT}) have not the
same status as the sum rules (\ref{sum2}), (\ref{sum4}) which are the
{\it theorems} of $QCD$.

Let us explain (very sketchy) how the results for the microscopic spectral
density have been derived.
For simplicity, assume that $\nu = 0$ so that $T$ is a square matrix:
 $n_L = n_R \equiv n$.
$P(T)$ and the Dirac determinant depend only
on the eigenvalues $\lambda_n$ of the matrix $T$. Therefore, it is convenient
to write the integral over ${\cal D}T$ as the integral over eigenvalues and do
the
integral over remaining angular variables on which the integrand does not
depend. We have
\footnote {This transformation and the whole stochastic matrix technique
is well known for nuclear physicists and solid state physicists who are
concerned with the problem of quantum dynamic for stochastic hamiltonia.
The particular problem studied here coincides with the known problem
of "Gaussian unitary ensemble". The vanishing of the Jacobian at $\lambda_k =
\lambda_l$ reflects the notorious repulsion of the levels. For an extensive
review of the relevant mathematics see \cite{Mehta}.}
  \be
  \label{Jacob}
{\cal D}T = C\prod_{k<l} (\lambda_k^2 - \lambda_l^2)^2 \prod_k \lambda_k
d\lambda_k
  \ee
Thus, joint eigenvalue distribution is just
  \be
  \label{rovse}
\rho(\lambda_1,\ldots, \lambda_n) \sim
\prod_{k<l} (\lambda_k^2 - \lambda_l^2)^2 \prod_k \lambda_k \
\exp \{ - \frac An \sum_{k=1}^n \lambda_k^2 \} \prod_f
(\lambda_k^2 + m_f^2)
  \ee
The microscopic spectral density is obtained after integration of (\ref{rovse})
 over all eigenvalues but one. After some work, one gets the answer in the
limit $n \rightarrow \infty$. In \cite{Izm} the result was obtained in the
particular case $\nu = 0$. Being expressed via physical variables
$V$ (the volume of the system) and $\Sigma$ (the quark condensate),
$A \equiv (\Sigma V)^2/4$, it has the form
  \be
  \label{romic}
\rho(\lambda) = \frac {\Sigma^2 V\lambda}2 \left[
J^2_{N_f} (\Sigma V \lambda) - J_{N_f+1}(\Sigma V \lambda) J_{N_f-1}(\Sigma
V\lambda) \right]
  \ee
where $J_\mu(x)$ are the Bessel functions.
The result (\ref{romic}) has been obtained in \cite{Izm} under the assumption
$\nu = 0$. In \cite{nuJac} it has been generalized to the case $\nu \neq 0$.
[the only change is that $N_f$ should be substituted by $N_f + |\nu|$ as in
Eqs. (\ref{sum2},\ \ref{sum4})].

At very small $\lambda \ll
1/V\Sigma $, the spectral density is suppressed $\rho(\lambda) \sim
\lambda^{2N_f+1}$ (the suppression is due to the determinant factor
(\ref{detNf}) which punishes small eigenvalues). But then it rises
and after several oscillations with decreasing amplitude levels off at
a constant $\rho(0) = \Sigma/\pi$ as it should in the thermodynamic limit
$V \rightarrow \infty$. The function (\ref{romic}) with $N_f = 0,1,2$ together
with the results of numerical instanton calculations are
plotted in Fig.3 borrowed  from Ref.\cite{JacPol}.
Likewise, one can integrate Eq.(\ref{rovse}) over all eigenvalues but
two and derive the expression for the correlation function $\rho(\lambda_1,
\lambda_2)$ etc.

The sum rules derived in the previous section can be expressed as the integrals
of $\rho(\lambda), \rho(\lambda_1, \lambda_2)$ etc. with a proper weight.
For example, the simplest sum rule is
   \be
  \left< \sum_{\lambda_n > 0} \frac 1{\lambda_n^2} \right>_{\nu = 0}
= \int_0^\infty \frac {\rho(\lambda) d\lambda}{\lambda^2} =
\frac {V^2\Sigma^2}{4N_f}
   \ee
which coincides with (\ref{sum2}). Thus, the excellent agreement of the model
instanton calculations with the stochastic matrix model results displayed
in Fig.3 implies also the agreement of the former with the exact theoretical
results (\ref{sum2}), (\ref{sum4}) etc.

 \section{Exotic theories.}
\setcounter{equation}0
The same program as in $QCD$ can be carried out in other gauge theories
with non-standard fermion content. The results derived earlier depended
on the assumption of the standard pattern of chiral symmetry breaking
  \be
  \label{chif}
SU_L(N_f) \otimes SU_R(N_f) \rightarrow SU_V(N_f)
   \ee
so that $N_f^2 - 1$ Goldstone bosons living on the coset, which is also
$SU(N_f)$, appear. However, the breaking according to this scheme only occurs
for fermions belonging to the complex (e.g. fundamental) representation of the
gauge group. The $SU(N_c)$ group involves truly complex representations
 when $N_c \geq 3$. For $N_c = 2$, the fundamental
representation (and also other representations with half-integer isospin)
is pseudoreal: quarks and antiquarks transform in the same way
under the action of the gauge group and the pattern of chiral symmetry breaking
{\it is} different leading to different sum rules. A third pattern of chiral
symmetry breaking is for fermions in the real (e.g. adjoint)
representation leading to yet another class of sum rules.

Consider first the case of pseudoreal fermions. It may have a considerable
practical importance as the numerical calculations with the $SU(2)$ gauge
group are simpler than with $SU(3)$ group, and it may be easier to confront
the sum rules with lattice simulations. The true chiral symmetry group of the
$QCD$-like lagrangian but with $SU(2)$ gauge group is $SU(2N_f)$ rather
than just $SU_L(N_f) \otimes SU(N_R) \otimes U_V(1)$ (it involves also
the transformations that mix quarks with antiquarks). The pattern of
 spontaneous symmetry breaking due to formation of quark condensate is
\cite{Peskin}
  \be
  \label{chif2}
SU(2N_f) \rightarrow Sp(2N_f)
  \ee
The breaking (\ref{chif2}) leads to the appearance of
$$(2N_f)^2 - 1 - (2N_f^2 + N_f) = 2N_f^2 - N_f - 1 $$
Goldstone bosons which are parametrized by the coset $SU(2N_f)/Sp(2N_f)$.
For $N_f = 2$ , we have 5 instead  of the usual 3 Goldstone bosons. Let
us now allow for small non-zero quark masses (which break the chiral
symmetry explicitly and give non-zero masses to Goldstone particles) and
study the dependence of the partition function in the sector with a given
topological charge $\nu$ on the quark mass matrix ${\cal M}$ and the volume
$V$. The analog of the result (\ref{Znu}) for the pseudoreal case which
is valid in the region (\ref{region}) is the following (see \cite{jaJac})
for more details)
  \be
  \label{Znu2}
Z_\nu \sim \int_{U(2N_f)} d\mu(U) ({\rm det} U)^{-\nu}
\exp \left\{\frac {V\Sigma}2 {\rm Re Tr}
\{{\cal M}UIU^T\} \right\}
  \ee
where
  \be
  \label{I}
I = \left( \begin{array}{cc} 0&{\hat 1} \\ {- \hat 1}&0 \end{array} \right)
  \ee
is the symplectic $2N_f \otimes 2N_f$ matrix. Formally, the representation
(\ref{Znu2})
involves the integral over $U(2N_f)$, but actually it is the integral
over the coset $SU(2N_f)/Sp(2N_f)$ involving less number of parameters
({\it and} the integral over the $U(1)$ part of $U(2N_f)$ which is nothing
else but a vacuum angle $\theta$) --- multiplying $U$ by a simplectic
matrix $\in Sp(2N_f)$ leaves the integrand invariant.

Expanding (\ref{Znu2}) over the group invariants involving the quark matrix
${\cal M}$ and expanding also the quark-gluon representation of the same
partition function, the sum rules for the inverse powers of the eigenvalues
can be derived. The simplest sum rule is
   \be
  \label{sum22}
\left< \sum_{\lambda_n > 0} \frac 1{\lambda_n^2} \right>_\nu = \frac
{V^2\Sigma^2}{4(|\nu| + 2N_f - 1)}
  \ee

Consider now the third non-trivial class of theories where the fermions
belong to the real representation of the gauge group. The supersymmetric
Yang-Mills theories involving Majorana fermions in the adjoint color
representation belong to this class. We will not assume, however,
 that the theory is supersymmetric and consider a Yang-Mills theory
coupled to $N_f$ different Majorana adjoint fermion fields. Also, we
will assume that the fermion condensate is formed and the chiral symmetry
which the lagrangian enjoys for $N_f \geq 2$ is spontaneously broken
\footnote{ This  dynamic
assumption is not so innocent. For example, in $N=2$ supersymmetric
 Yang-Mills theory it does not happen \cite{seiberg}. But this theory
 involves also massless adjoint scalar fields with  particular interaction
vertices. It is quite conceivable that the non-supersymmetric theory
with only gauge fields and two Majorana adjoint flavors {\it involves}
chiral symmetry breaking. All the results are derived under this assumption.}.

The chiral symmetry group of such a theory is $SU(N_f)$. Formation of the
condensate breaks it down to $SO(N_f)$, and
$$ N_f^2 - 1 - \frac {N_f(N_f -1 )}2 = \frac {N_f(N_f+1)}2 - 1 $$
Goldstone bosons appear. Again, the partition function $Z_\nu$ is given by
the integral over the coset $SU(N_f)/SO(N_f)$ which has the form
  \be
  \label{Znuad}
Z_\nu \sim \int_{U(N_f)} d\mu (U) ({\rm det} U)^{-2\nu N_c}
\exp \left\{V\Sigma\ {\rm Re\ Tr}
\{{\cal M}UU^T\} \right\}
  \ee
The essential difference with the case of fundamental fermions
 is that the admissible values for the topological charge $\nu$ are not
integer but integer multiples of $1/N_c$. That means, in particular, that
the function $Z({\cal M}, \theta)$ given by the Fourier sum (\ref{ZFour})
is a periodic function of $\theta$ with the period $2\pi N_c$ (not $2\pi$
as before). See Ref.\cite{Leut} for a detailed discussion of this important
issue.

To derive the sum rules, we have to expand (\ref{Znuad}) over ${\cal M}$
and to compare it with the expansion of the quark-gluon path integral for
the partition function. And here we meet a known problem characteristic
for theories with Majorana particles. The matter is that the reality condition
for the Fermi fields can be fulfilled in Minkowski space but not in Euclidean
space - it immediately leads to a contradiction \cite{Ramond}

However, the Euclidean path integral for the partition function in a theory
with Majorana fermions still can be defined by analytic continuation
of the Minkowski space path integral. In the Minkowski space, the path
integral over fermion fields is equal to the {\it square root} of the
 corresponding Dirac determinant. The latter can be continued to Euclidean
space without problems, and the square root is also taken easily here due
to a notable fact that the spectrum of the Euclidean Dirac operator for the
adjoint fermions is doubly degenerate: for any eigenfunction $u_n(x)$
with an eigenvalue $\lambda_n$, the function $C^{-1}u_n^*(x)$ ($C$ is the
charge conjugation matrix) is linearly independent from $u_n(x)$ and has
exactly the same eigenvalue \cite{Leut}. Thus, square root of the Dirac
determinant is just
  \be
  \label{detMaj}
[{\rm det} (-i\not\!\!{\cal D} + \tilde{{\cal M}}]^{1/2} \ = \
(det_f {\cal M})^{\nu N_c} \prod_{\lambda_n > 0}'
det_f (\lambda_n^2 + {\cal M} {\cal M}^\dagger)
  \ee
where $\tilde{{\cal M}} = \frac 12 (1-\gamma^5) {\cal M} + \frac 12
(1+\gamma^5) {\cal M}^\dagger$,
the product counts only one eigenvalue of each degenerate pair, and the
factor $(det_f {\cal M})^{\nu N_c}$ reflects the presence of $\nu N_c$ pairs
of zero eigenmodes of adjoint Dirac operator in the gauge field background
with the topological charge $\nu$
\footnote{The nice result (\ref{detMaj}) is specific for the theories with
{\it adjoint} Majorana fermions. To define in the Euclidean space a theory
like the standard model
involving chiral fermions in the fundamental representation is more difficult
\cite{VZ}.}.

The final form of the simplest sum rule for the eigenvalues of the {\it Dirac}
adjoint determinant is
     \be
  \label{sum2ad}
\left< \sum_{\lambda_n > 0}' \frac 1{\lambda_n^2} \right>_\nu =
\frac {V^2\Sigma^2}{4(|\nu|N_c + (N_f + 1)/2)}
  \ee
where, again, only one eigenvalue of each degenerate pair is taken into account
in the sum.

The sum rules  (\ref{sum2}), (\ref{sum22}), and (\ref{sum2ad}) can be written
universally as
   \be
  \label{sum2u}
\left< \sum_{\lambda_n> 0} \frac 1{\lambda_n^2}\right>_\nu = \frac
{V^2\Sigma^2}
{4\{|\nu| + [{\rm dim\ (coset)} +1]/N_f\}}
  \ee
with the rescaling $\nu \rightarrow \nu N_c$ and counting in the sum only one
eigenvalue of each degenerate pair in the adjoint case.

The sum rules   (\ref{sum22}), and (\ref{sum2ad}) and their analogs with
higher inverse powers of $\lambda_n$ can be derived also in the stochastic
matrix technique, and not only the sum rules, but also the expressions for
the microscopic spectral density $\rho(\lambda)$, correlators $\rho(\lambda_1,
\lambda_2)$  etc. \cite{3way}

Mathematically,  3 classes of theories discussed (with complex fermions,
pseudoreal fermions and real fermions) are described in terms of 3 classic
stochastic ensembles: Gaussian unitary, Gaussian orthogonal, and Gaussian
simplectic.
It is noteworthy that, though the sum rules in all 3 cases are the same
 for $N_f =1$ [where no spontaneous breaking of chiral symmetry occurs, no
Goldstone bosons
appear, and the partition function has the extensive form (\ref{Zext})] , the
microscopic spectral densities and the correlators are not.
That elucidates again the fact that , when deriving microscopic spectral
densities in the stochastic matrix technique, an extra assumption (\ref{PT})
beyond using just the symmetry properties of the theory is neccessary to adopt.

\section{Acknowledgements.} It is a pleasure for me to thank the organizers
of the Seoul meeting for kind hospitality. I am indebted to J. Verbaarschot
for valuable discussions and for kindly sending me the postscript files
of Fig.2 and Fig.3. This work was supported in part by Schweizerisher
Nationalfonds and the INTAS grant 93-0283.

\section*{Figure captions.}
{\bf Fig. 1}. 2-point correlator. (a) Quark representation. The loop is
evaluated in a gauge-field background to be averaged over afterwards.
(b) Quasi-Goldstone contribution.

\vskip 0.3cm
\noindent {\bf Fig. 2}. Instanton calculations for the spectral density
$n(\lambda) = \rho(\lambda)$ for different $N_f$. $\lambda$ is measured in
units of $\Lambda_{QCD}$, and the area below the curve is normalized to 1.

\vskip 0.3cm
\noindent {\bf Fig. 3}. Microscopic spectral density as a function of
$z = \Sigma V \lambda/\pi$. The normalization $\rho(0) =1$ is chosen.
Dashed lines correspond to
the theoretical result (\ref{romic}). Full lines present the results of
numerical calculations in the instanton liquid model.

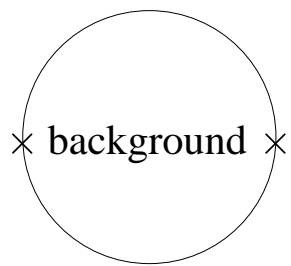
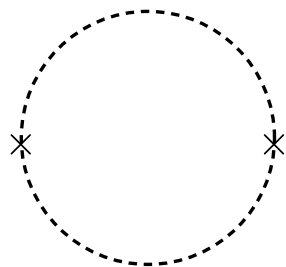
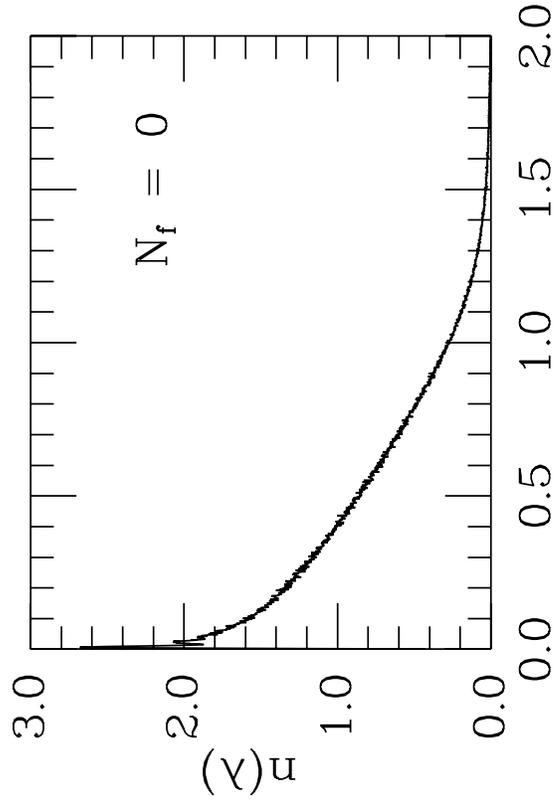
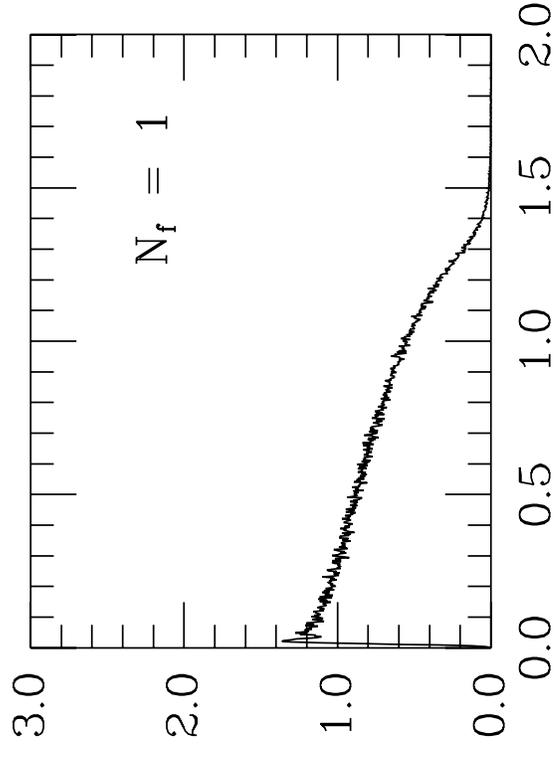
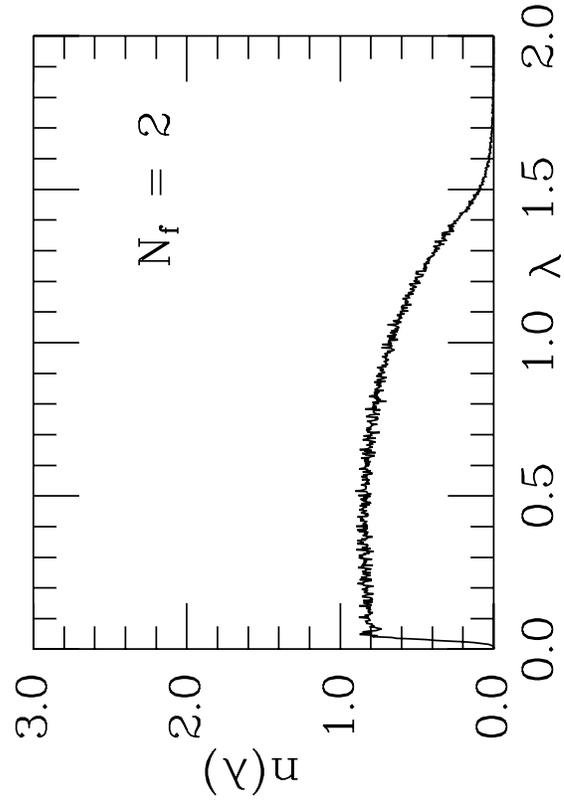
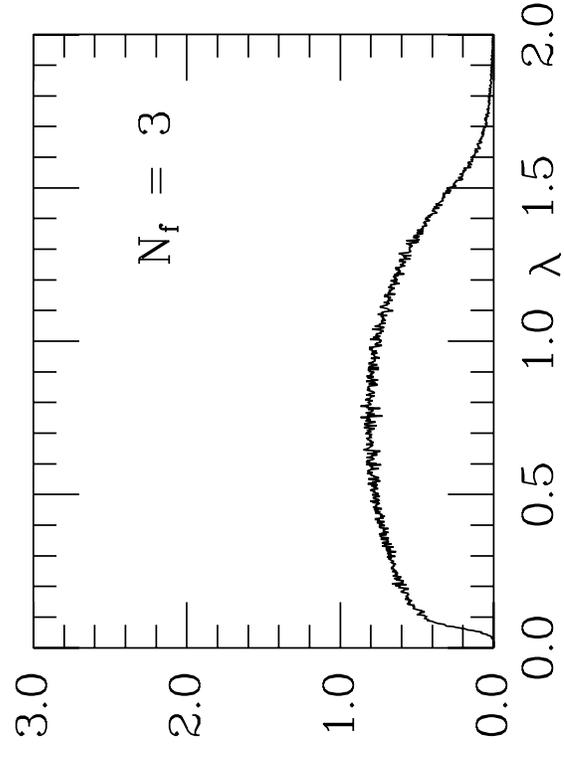
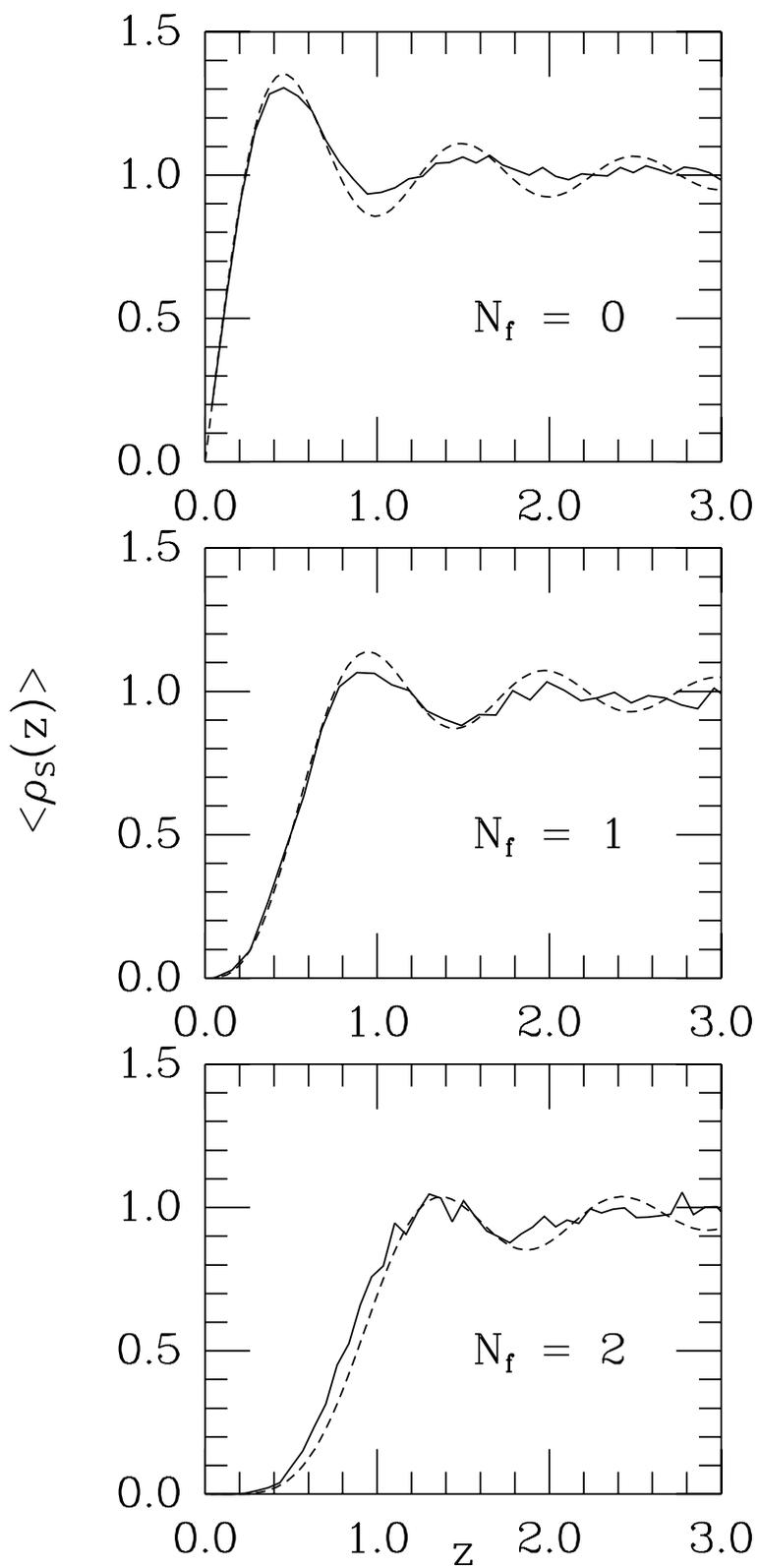
\end{document}